# Capturing Information Flows inside Android and Qemu Environments

Marco Sironi, Francesco Tisato

Università degli Studi di Milano – Bicocca
Dipartimento di Informatica, Sistemistica e Comunicazione
Viale Sarca 336, 20126 Milano – Italy
`marco.sironi@disco.unimib.it`
`francesco.tisato@disco.unimib.it`

February 12, 2013

**Abstract –** The smartphone market has grown so wide that it assumed a strategic relevance. Today the most common smartphone OSs are Google's Android and Apple's iOS. The former is particularly interesting due to its open source nature, that allows everyone to deeply inspect every aspect of the OS. Android source code is also bundled with an hardware emulator, based on the open source software Qemu, that allows the user to run the Android OS without the need of a physical device. We first present a procedure to extract information flows from a generic system. We then focus on Android and Qemu architectures and their logging infrastructures. Finally, we detail what happens inside an Android device in a particular scenario: the system boot.

## 1 Introduction

It is known that the number of people that have access to the Internet is constantly increasing, but it is particularly interesting to have a look at how they do it. According to ITU [1], at the end of 2011 17% of global population accessed the Web through a mobile device and 45% was covered by a 3G mobile network. StatCounter [2] shows that in October 2012 12.3% of web pages were hit from a mobile device and that this percentage has been always increasing in the last year. According to The NPD Group [3], total smartphone sales rose 9% in the second quarter of 2012 compared to the same period of previous year.

These numbers show clearly that the mobile devices market, and in particular smartphones' one, is constantly growing larger, assuming an even more strategic relevance. In this scenario the ability of inspecting the internal mechanisms of a device becomes crucial for different figures: developers who want to write efficient applications, engineers who want to improve the system's performances or remove bugs, etc. For this reason we will show a procedure to capture the internal information flows of a mobile device.

Actually the smartphone scene is dominated by two OSs: Apple's iOS and Google's Android [4]. In this paper we concentrate our attention on Android, due to its diffusion and to its *open source* nature. In fact, the availability of the Android source code allows everyone to deeply inspect its internals and to trace system information at any level of detail.

A smartphone is not only made up of a software layer, but also of the underlying hardware platform. In some cases it can be necessary to collect information from this component, but this will require to physically insert some kind of probe in it. This operation is definitely not within the reach of everyone: who wants to perform it must have the adequate instrumentation (the probe itself, a welding machine, etc.) and skills (must know where to attach the probe, how to do it and how to get the data from it). Luckily, Google provides an hardware emulator, based on the open source software Qemu. This means that we can easily put software probes directly into its source code, obtaining the hardware information in an easy and cheap way. So in this paper we will not only show how to capture information flows from the software running on a mobile device, but also how to get them from its (emulated) hardware.

The remainder of this paper is structured as follows. In Section 2 we present a general procedure to retrieve information flows from a generic system. In Section 3 we introduce the Android environment, its structure and a set of prebuilt instruments that can be used to extract information flows. In Section 4 we present the Qemu platform, its architecture and its built-in logging infrastructure. In this section we will also focus on customizations made by Google to the standard version of Qemu. In Section 5 we study a particular scenario: the boot phase of a device running Android; this can come handy in case we need to run a software probe on system start-up.

## 2  A general procedure

We can get a great variety of data from a device. For this reason the first thing we must state is *why* we want to capture information flows. Depending on what our goal is, we will need very different kind of information. For example, consider the case where an application crashes and we want to perform some debugging. We are interested in fine-grained, deterministic information about single executions, such as the execution state, parameters, variable values and so on. Now consider another case, where we want to optimize an application. Our goal is to understand what the most time-consuming methods are, so that we can analyze and eventually edit them to make the application faster (this operation is usually called "application profiling"). This time we are looking for a completely different kind of information: we want to know what methods of the application are called, how often they are called and how long each of them takes to be executed. In this case we are mostly interested in statistical information obtained aggregating data coming from an high number of executions. Another important aspect we must care about when we trace a device is to understand what are the *abstraction layers* we need to inspect. For example we must understand if we need information at application methods level, at system call level of at physical operation level detail. A device, as well as a generic system, is usually too complex to be represented in any of its details; for this reason abstraction layers are used to highlight different aspect of the system that may be relevant, hiding all others. It is fundamental to understand what abstraction levels show the information we need and focus on them. It is often necessary to capture information flows from different abstraction layers and then determine what are the relations between them.

Once our goal is clear, the next step is to state *what* information we need to reach it. We must collect *all necessary information* and *nothing else*, because any unnecessary information could lead to unwanted side effects. In particular two problems could arise: *time overhead* and *information overflow*. We have time overhead when information catching requires too much time and the application or the system we want to trace is slowed down by our probes. Time overhead is particularly relevant when we need to catch time-accurate data. Notice that when we insert any kind of probe into a system we *always* generate time overhead, but in many cases the overhead produced by the instrumentation is so low that it can be neglected. Information overflow happens when our instrumentation uses a big amount of the system memory (either RAM or disk space). We must care of this problem when we want to trace disk or memory usage, because we will need a mechanism to discriminate the amount of memory used by our probes by the one actually used by the application or system we are tracing. Another scenario where we must care about information overflow is tracing systems with low memory resources. Also notice that information overflow can lead to time overhead: for example, if we use an instrumentation that makes heavy use of RAM it may happen that the system must frequently use the swap file (if present), slowing down significantly.

When we know exactly what information we need, we must deal with the *structure* of the system we want to trace. In fact, we must understand *where*, in this structure, information of our interest is located.

Once we know where to find the data we need, it is necessary to decide *how* to extract them from the system we are analyzing. In this phase, it is extremely important to understand if the system we are tracing already provides a logging infrastructure that can be used to obtain the data we need. Skipping this operation can easily lead to "reinvent the wheel", i.e. to insert external probes that are not necessary at all. External probes should only be inserted once stated that information of our interest cannot be extracted using existing methods. Complex systems usually provide different logging infrastructures, each designed for a particular purpose. Note that a particular logging infrastructure may not allow to obtain the data we are looking for, while another may do. Then we need to perform two further operations: first, we must choose a *format* to store the data; second, we must decide where to put all information we capture from the system.

The choice of the format can significantly affect time overhead and information overflow: a compact format could cause time overhead, because the CPU must compress the data we are intercepting; on the other side, a verbose format could lead to information overflow. The format is strictly related to the software we intend to use to process the data we get from the system: the most immediate choice is to save our information flows in one of the formats supported by the application we will use to process them; however, this could lead to time overhead or information overflow. In this case we can use a data format that grants acceptable time overhead and information overflow and then perform an "offline" conversion in a software-readable format at a later time.

The last thing we must do is to decide where to store the captured data. We substantially have two options: save information on the same system we want to trace or save them in an external location. In the first case we shall find a way to transfer the information flows from the traced system to the one where the data will be processed (obviously if we want to process the captured data on the same system we traced this operation is not needed). In the second case we can store the captured data either on an external memory, that will be used to transfer the data on the system we will use to process them at a later time, or directly on a second system. In the latter case we will need a reliable communication channel between the two systems.

## 3 Information flows in Android

Android [5] is an open-source software stack for mobile phones and other devices, developed and maintained by the Android Open Source Project (AOSP), a consortium of technology and telecommunication companies led by Google. The majority of the Android software is licensed with Apache Software License 2.0 and the instructions to download all its source code can be found on the AOSP website. In this section we will explain the overall Android architecture and then show different prebuilt tools that can be used to extract information from this system.

### 3.1 Android architecture

Figure 1 (source: [6]) shows the major components of the Android platform. Blue components are written in Java, while green ones are written in C/C++. At the lowest level of the system we find the *Linux kernel*. It is necessary to make a clarification: Android is built on the Linux kernel, but Android *is not* Linux. For example it does not include all standard Linux utilities, it does not include a native windowing system and it does not support the GNU C Library (libc). The first version of Android relied on a standard Linux 2.6.24 kernel patched to enhance some aspects such as power management (a critical feature for mobile devices) and inter-process communication mechanism [7]. Kernel contains different drivers that allow it to communicate with hardware peripherals (e.g. display, keypad, camera, etc.).

Above kernel we find a set of *C/C++ (native) libraries* used by different components of Android. In particular:

- "bionic" libc is a BSD-derived implementation of the GNU C Library, customized for embedded devices;
- Media Libraries support playback and recording different audio, video and image formats;
- Surface Manager manages access to the display subsystem and composites 2D and 3D graphic layers from multiple applications;

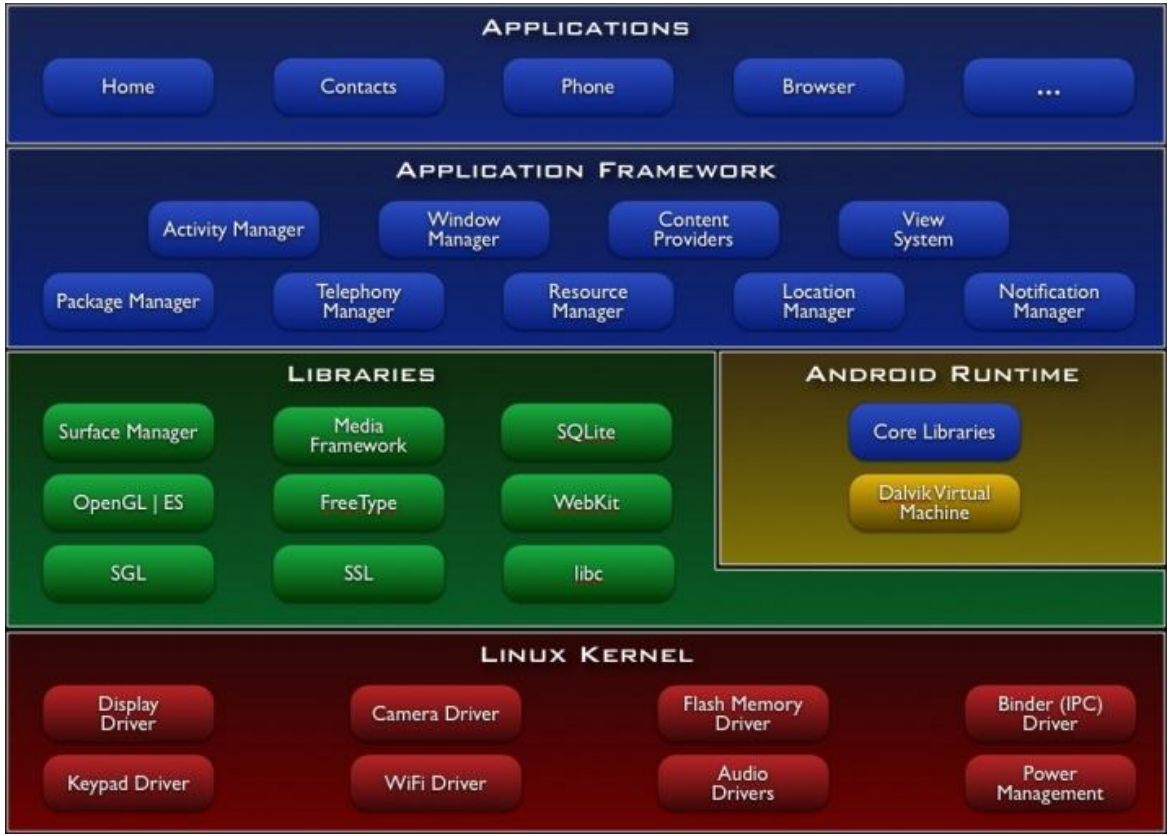

**Figure 1 – Android architecture overview (source: "App Framework | Android Developers" [6])**

- LibWebCore is a modern web browser engine;
- SGL is a 2D graphics engine;
- 3D libraries are based on OpenGL ES 1.0 APIs and use either hardware 3D acceleration or the included 3D software rasterizer;
- FreeType is a library for bitmap and vector font rendering;
- SQLite is a lightweight relational database engine available to all applications.

A particular native library is the Hardware Abstraction Layer (HAL) [7], [8]. Unlike standard Linux, Android does not only require proper device drivers to function on hardware. HAL is a user-space C/C++ library layer that separates the Android platform logic from the hardware interface: it provides services to the upper layer, the Android application framework, and at the same time it uses functions provided by the lower layer Linux device drivers. HAL defines an API for each type of hardware supported by Android's core. In order for a specific hardware to properly work with Android, it must provide a hardware "module" (not a kernel module!) which conforms to the API specified for that type of hardware. On standard Linux systems the kernel itself acts as an abstraction layer between the hardware platform and the user-space, providing interfaces that applications can use to access hardware devices. So, why does Android need a further abstraction layer? The answer is: to let the upper levels of Android be more kernel-independent as possible.

At the same level of the C/C++ libraries we find the *Android runtime*. Android runtime is made of two components: a set of libraries, that provides the core API for the Java language, and the Dalvik Virtual Machine. The Dalvik VM has been specifically designed for embedded environments, where resources like power, CPU, memory size and storage size are limited. Every Android application runs in its own process, with its own instance of the Dalvik virtual machine. That is important for different reasons [9]: first, no application is dependent upon another; second, if an application crashes, it shouldn't affect any other running application; third, it simplifies memory management. The Dalvik VM executables use the Dalvik Executable (.dex) format, which is optimized for minimal memory footprint. Android developers can write applications in Java and then compile them; the resulting executables can easily be converted into the .dex format by the "`dx`" tool, shipped with Android source code (and also Android SDK [10]). The Dalvik VM relies on the Linux kernel for underlying functionality such as threading and low-level memory management.

Above native libraries and Android runtime we find the a*pplication framework*, that provides essential services to the Android platform. For example, Content Providers enable applications to access data from other applications or to share their own data; Resource Manager provides access to non-code resources (localized strings, graphics, layout files, etc.) and Notification Manager enables all applications to display custom alerts in the status bar.

At the highest level of Android architecture we find a set of core a*pplications*, all written in Java. As already said, each Android application runs in its own sandboxed Dalvik VM.

### 3.2  Android logging system

Figure 2 (source: [11]) shows an overview of the Android logging infrastructure. It allows to separate Android applications logging information from Linux kernel one (which is stored using the `printk` function in the kernel and accessed using `dmesg` or `proc/kmesg`). However, we will see that Android logging system stores messages in kernel buffers.

The Android logging system consists of:

1. A Java class, `Android.util.Log`, that allows Android applications to write log messages;
2. A native library, `liblog`, that allows native programs to write log messages;
3. A kernel driver, `logger`, and four circular kernel buffers that allow to store log messages;
4. A stand-alone tool, `logcat`, that allows to read log messages from buffers directly on the device;
5. A device-side program, `adbd`, and a host-side program, `adbserver`, that allow to view and filter log messages from the host machine (to which the device is attached).

`Android.util.Log` can be included in any Java program and provides static methods to write log messages with a tag and a priority. Tag and priority can be used to filter log messages. Priority levels are (from lower to higher): VERBOSE, DEBUG, INFO, WARN, ERROR. It is important to note that by default Android redirects standard output and standard error to `/dev/null`; this means that anything printed invoking `System.out.println()` or `System.out.println()` will be lost. However, it is still possible to reroute anything destined to `System.out` and `System.err` to kernel buffers used to store log messages with some additional configuration [12].

The native library `liblog` can be used by any user-space C/C++ application to save log messages in kernel buffers and is also invoked by the `Android.util.Log` class to store log messages coming from Java applications in kernel buffers. As already said for Java applications, standard output and standard error are rerouted to `/dev/null`, but a tool, `logwrapper`, can be used to run a program and redirect it's standard output into log messages.

The `logger` kernel driver supports four circular buffers:

- *main*, the main application log;
- *events*, for system event information;
- *radio*, for radio and phone-related information;
- *system*, for low-level system messages and debugging (only in latest versions of Android).

Log reading and writing is done via normal Linux file reads and writes. The write path is optimized, so that an `open() – write() – close()` sequence can occur very quickly, to avoid logging having too much time overhead in the system [13].

The `logcat` tool reads and filters log messages stored in the abovementioned kernel buffers. By default it reads the *main* log buffer. It can be run directly on the Android device or on the host machine via the Android Debug Bridge (ADB) [14][15][16]. ABD is a client-server program made of three components:

- a client, which runs on the host machine;
- a server (`adbserver`), which runs as a background process the host machine and manages communication between the client and the ADB daemon running on the Android device;
- a daemon (`adbd`), which runs as a background process the Android device.

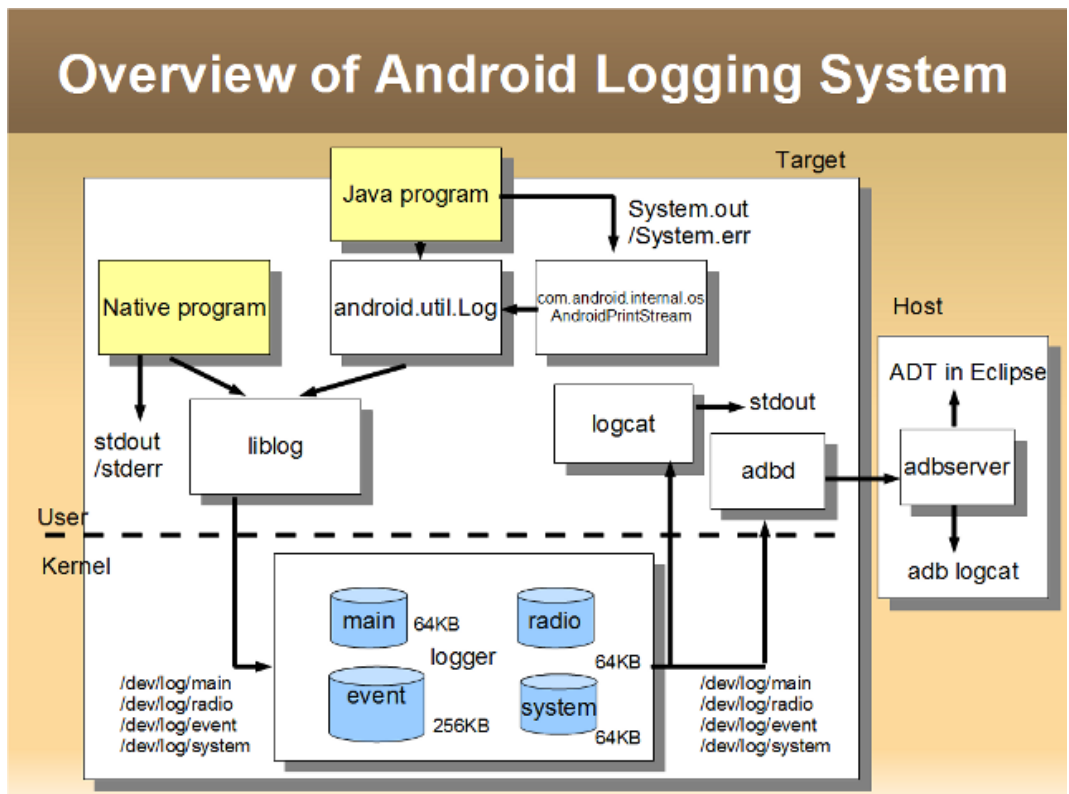

**Figure 2 - Android logging system overview (source: T. Kobayashi, "Logging system of Android" [11])**

When an ADB client is started, it first checks whether there is an ADB server process already running and if there is not, it starts the server process. When the server starts, it binds to local TCP port 5037 and listens for commands sent from ADB clients (communication between ADB clients and server is performed on this port). The server then sets up connections to all running devices. Android devices are located by scanning odd-numbered ports in the range 5555 to 5585. Where the server finds an ADB daemon, it sets up a connection to that port. Note that each device acquires a pair of sequential ports – an even-numbered port for console connections and an odd-numbered port for ADB connections. Once the server has set up connections to all devices, ADB clients can be used to send commands to them. Because the server manages connections to emulator/device instances and handles commands from multiple ADB clients, any device can be controlled from any client.

## 3.3 Some useful tools

Android source code includes the *Device Monitor*, a stand-alone program that groups different debugging and analysis tools that can be used to extract information from a mobile device. The included tools are: Dalvik Debug Monitor Server (DDMS), Tracer for OpenGL ES, Hierarchy Viewer and Traceview.

### 3.3.1 Dalvik Debug Monitor Server

DDMS [17] is a powerful debugging tool that provides a lot of features, such as thread and heap information on the device, log messages reading, process and radio state information, incoming call and SMS spoofing and location data spoofing.

As already said, on Android every application runs in its own process, each of which runs in its own virtual machine. Each VM exposes a unique port that a debugger can attach to. When DDMS starts, it uses ADB to establish a connection to each VM's debugger. DDMS talks to the VMs using a custom communication protocol.

DDMS allows to easily track heap usage of each process (i.e. application) running on the device and to show all running threads of a specific process. Another interesting feature is the capability to track objects that are being allocated to memory and to see which classes and threads are allocating them. This information is useful for assessing memory usage that can affect application performance.

DDMS integrates Traceview (described in Section 3.3.4), a tool that allows to perform method profiling. One last interesting aspect of DDMS is that it integrates `logcat`, so it allows to read log kernel buffers without the need to explicitly launch the `logcat` command via ADB.

### 3.3.2 Tracer for OpenGL ES

Tracer is a tool that allows to analyze OpenGL for Embedded System code in Android applications. It captures OpenGL ES command execution logs and can also capture progressive images of the frames generated by those commands. These data can be used to perform logical and visual analysis of the code. Tracer is only working with devices running Android 4.1 or higher.

### 3.3.3 Hierarchy Viewer

Hierarchy Viewer is a tool for user interfaces (UIs) debugging and optimization. It provides a visual representation of the layout's View hierarchy (a View is the basic block for user interfaces) and a magnified inspector of the display. For each View object, the Hierarchy Viewer displays rendering performance data, allowing to detect bugs and to highlight parts of the UI with render performances so slow that can be a problem.

### 3.3.4 Traceview

Traceview is a graphical viewer to see logs created by an Android application. These logs contain information needed to perform application profiling, i.e. to track metrics such as number of calls to a method, a method's execution time and time spent executing a method. Developers can add trace code inside their applications and then use Traceview to view the recorded execution information. As already said, Traceview can also be launched directly from DDMS: this method is less precise than the previous one, as the user can only choose when to start and end tracing. In other words, this method does not allow to know exactly what region of code has been executed during tracing.

# 4 Information flows in Qemu

Qemu [18], [19] is a generic and open source machine emulator and virtualizer, originally developed by Fabrice Bellard and first released in 2003. Emulation and virtualization are two different concepts that are often confused [20]. Emulate a machine means to simulate all its hardware in software; this allows to run OSs and applications for a given architecture *A* on a machine with a totally different architecture *B*. For example, an emulator like Qemu allows to run OSs and applications developed for an ARM machine (guest) on an x86 machine (host). Virtualize a machine means to simulate only parts of its hardware, while most operations are still executed on real hardware. A software module called *virtualizer* is responsible for intercepting operations coming from the virtual machine that must be sent to the "fake" hardware parts so that the guest remain unaware it is not running on real hardware. Virtualization is usually faster than emulation, but it can only be used when the real hardware has the same architecture of the guest machine.

QEMU runs on x86 systems running Linux, Microsoft Windows and some UNIX platforms, and can host target systems from a range of different microprocessors [21]. Most of the smartphones actually on the market is ARM-based [22], so we are interested in using Qemu as an emulator, not as a virtualizer.

## 4.1 Qemu architecture

Qemu is made of several components:

- CPU emulator, that emulates architectures such as x86, PowerPC, ARM, SPARC, etc.;
- Emulated devices such as VGA display, serial port, PS/2 mouse and keyboard, IDE hard disk, network card, etc.;
- Generic devices (e.g. block devices, character devices, network devices) that are used to connect the emulated devices to the corresponding host devices;

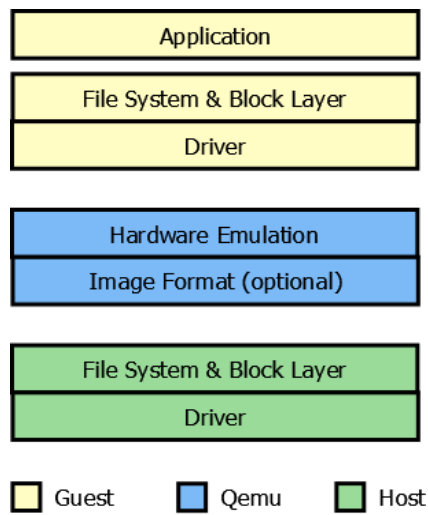

**Figure 3 – Qemu storage overview (source: S. Hajnoczi, "An Updated Overview of the QEMU Storage Stack" [25])**

- Machine descriptions (e.g. PC, PowerMac, Sun4m) that instantiate the emulated devices;
- Debugger;
- User interface that allows to interact with the emulated devices.

To provide CPU emulation Qemu uses a mechanism called Tiny Code Generator (TCG) [23], [24], which allows runtime conversion of instructions from guest CPU to host CPU. This operation is known as dynamic binary translation or Just-in-Time (JIT) compilation. Qemu first translates native assembly code of the guest processor, for example ARM, into a sequence of *micro-operations*. In a second step, these micro-operations are converted in executable code for the host processor, for example x86, and packaged in *translation blocks*. These translation blocks are held in a cache memory and can be used again.

In the wide variety of devices Qemu can emulate, storage devices are particularly interesting. Qemu works with different kind of storage devices, including hard disks, floppy disks, CD/DVDs, USB memory sticks, on-board Flash memory and Secure Digital cards.

Qemu can either use real storage devices attached to the host machine or emulate them. When a storage device is emulated, its content must be stored in a disk image file on the host machine. Figure 3 (source: [25]) shows the Qemu storage stack, assuming it is working on a storage device image file. Guest application and kernel work exactly as they are running on real hardware. Guest talks to Qemu via an hardware emulation layer and Qemu performs I/O operations on the image file on behalf of the guest. Host treats guest I/O like any other user-space application.

Qemu supports several image types, including *raw* (a plain binary image of the storage device, very portable), *vmdk* (VMWare image format) and *vdi* (Oracle Virtualbox image format).

Qemu can also cache access to the disk image files and provides several methods to do that through the `cache` option, that controls the use of the cache for a specified drive. Supported caching mode are:

1. *Writeback*: will report data writes as completed as soon as the data is present in the host page cache. If the host crashes before the data is effectively written to the disk image, then the guest may experience data corruption.

2. *Writethrough:* the host page cache will be used to read and write data, but write notification will be sent to the guest only after Qemu has made sure to flush each write to the disk. This mode has a major impact on performance.

3. *None:* the host page cache will be completely avoided, so disk I/O will be performed directly to the guest page cache. The guest OS must handle the disk write cache correctly in order to avoid data corruption on host crashes.

4. *Directsync:* works like the *writethrough* mode with the difference that the host page cache will be avoided. Write notifications will be only sent to the guest when the data has been flushed to the disk.

5. *Unsafe:* this option tells Qemu that it never needs to write any data to the disk but can instead keep things in cache. Any problem occurring to the host can easily lead to data loss.

When Qemu is running, it provides a monitor console for interacting with Qemu. Through various commands, the monitor allows the user to inspect the running guest OS, change removable media and USB devices, take screenshots and audio grabs, and control various aspects of the virtual machine.

### 4.2 Qemu logging and tracing systems

Qemu provides two different infrastructures that can be used to extract information. The first one, called "logging" system, allows to view information related to CPU emulation, such as the generated host assembly code for each compiled translation block, the guest assembly code for each compiled translation block and the micro-operations for each compiled translation block. The second one, called "tracing" system [26], can be used to extract information for debugging, profiling, and observing execution of the guest system.

Qemu must be built with the appropriate option in order to provide the tracing infrastructure. In the Qemu source code a set of static trace events is defined in the `trace-events` file; each trace event contains the name of the event, its arguments and the format string which can be used for pretty-printing. The user can add custom trace events to the default ones. The trace events set is processed by the `tracetool` script during build to generate the `trace.h` header file. This file is then included by every source file that uses trace events.

The `tracetool` script automates trace event code generation and also keeps the trace event declarations independent of the trace backend. This way support for new trace backends can be added by extending the `tracetool` script. Actually the following trace backends are supported:

1. *Nop*: the trace events are disabled. This is the default backend;
2. *Stderr*: sends trace events directly to standard error, effectively turning trace events into debug `printf`s;
3. *Simple*: is shipped with the Qemu source tree, so it is portable. This is the recommended backend unless more advanced backends are specifically needed. It generates binary trace files that can be pretty-printed using the `simpletrace` script, that is also shipped with the Qemu code;
4. *Ust*: uses the LTTng Userspace Tracer library [27];
5. *Dtrace*: uses DTrace SDT probes but has only been tested with SystemTap [28].

Independently from the chosen backend, every time Qemu is launched users can select which of the events defined in `trace-events` must be traced.

### 4.3 Android Emulator vs. Qemu

The Android Emulator is shipped with the Android SDK and the Android source code, and is aimed to help developers in testing their applications without using a physical device. As previously said, the Android Emulator is based on Qemu, but *it is not Qemu*. In fact, Android Emulator has been built applying several customizations to standard Qemu code.

Unlike Qemu, the Android Emulator only emulates a specific platform, called Goldfish. The components of the Goldfish platform include ARMv5 CPU, MMC card and NAND flash storage device. Another major difference between Android Emulator and Qemu is that the former supports the *Android Debug Bridge*, that allows the user to easily interact with the emulator instances from the host machine. Android emulator also manages emulated devices in a different way compared to Qemu: using *Android Virtual Devices (AVDs)*. An AVD is an emulator configuration that lets the user model a device by defining hardware and software options to be emulated by the Android Emulator. This method simplifies the use of the emulator: the user needs to provide different system files and parameters to use Qemu, while he/she only has to specify a single parameter, the desired AVD, to run the Android Emulator. One last difference between the two software is that Android Emulator includes a *Memory Checker* subsystem, which is intended to catch simple memory access violations in the emulated system. To provide the Memory Checker subsystem with information about routine name, source file name and location for each detected memory access violation, the *ELFF library* has been introduced. The ELFF library is a full new implementation of an ELF/DWARF [29] parser.

## 5 Android boot sequence

Three components get involved in the boot phase of an Android system: the *bootloader*, the *kernel image* and the *root file system* (also called *initial RAM disk* or *initrd*) [30]. It is important to understand that the boot sequence of an Embedded Linux system (such as Android) is similar, but not equal, to a Standard Linux system's one.

Hardware configuration of Standard Linux machines typically include a board equipped with a non-volatile memory (ROM), where a program called BIOS is stored. This program gets executed by CPU after each reset; it performs basic hardware configuration and executes a second program, the *first stage bootloader*. The first stage bootloader is stored in the Master Boot Record, that usually corresponds to 512 byte located in the first sector of a mass storage device. Due to the lack of space, the first stage bootloader need to run another program, the *second stage bootloader*, to perform all operations typically required to a bootloader, i.e. kernel selection (in case more than one is available) and loading selected kernel image and optionally an *initial RAM disk* image. The initial RAM disk [31] is an initial root file system that is mounted prior to when the real root file system is available, and contains various executables and drivers that permit the real root file system to be mounted. If no initial RAM disk is provided, then the kernel image must have been compiled with all the modules needed to mount the real root file system. At this point the kernel is first uncompressed and then executed. During the boot of the kernel the initrd serves as a temporary root file system in RAM and allows the kernel to fully boot without having to mount any physical disk. After the kernel is booted, the real root file system is mounted and takes the place of the initial RAM disk, that is unmounted. After the kernel is booted, it starts the first user-space program, *init*.

Embedded Linux systems boot in a slightly different way [32]. In Standard Linux systems the BIOS, supplied by the hardware manufacturer, programs various core system components and provides various information for use by the OS, so the second stage bootloader only needs to concentrate on loading a kernel image (and possibly an initrd image) from a storage device and starting its execution.

Embedded systems typically don't come with handy, prewritten system vendor BIOS. For this reason, the system will need some of the heavy lifting to be done in the bootloader itself. The embedded bootloader will therefore need to initialize the RAM timings in the memory controller circuitry, flush the processor caches, and program up the CPU registers with sane default values. It will also need to determine precisely what hardware is installed within the system and supply this to the Linux kernel. On system start-up, Embedded Linux devices run a little piece of code, called *ROM boot code*, stored in an internal ROM. Unlike BIOS, ROM boot code only loads a *first stage bootloader* from a storage device (MMC, NAND flash, etc.) into an internal SRAM (DRAM has not been initialized yet). Due to limited SRAM size, the first stage bootloader only initializes DRAM and few other devices and then loads a *second stage bootloader* into DRAM. The second stage bootloader initializes other devices (network, USB, etc.) and loads the *kernel* and *initrd* images from storage device to DRAM, then it uncompress and runs the kernel. Unlike standard Linux machines, on embedded devices the initial RAM disk is usually not unmounted after the kernel is booted, but it serves as the permanent root file system. The first user-space program launched by the kernel is *init*.

Unlike other Linux systems (embedded or not), Android uses its own *init* program [33]. The Android *init* program processes two files and executes the commands it finds in them: *init.rc* and *init.<machine_name>.rc*, where *<machine_name>* is the name of the hardware that Android is running on. This usually is a code word; for example, "goldfish" is the name of the hardware emulated by the Android Emulator. The *init.rc* file is intended to provide the generic initialization instructions, while the *init.<machine_name>.rc* file is intended to provide the machine-specific initialization instructions. The Android *init* program performs two key operations: starts several daemons (adbd, usbd, etc.) and starts the *Zygote* process [34]. When the Zygote process starts, it initializes a Dalvik VM. Since every Android application counts on its own instance of the Dalvik VM to run in, VM instances are required to start quickly when a new application is launched. The Dalvik VM initialized by Zygote preloads and preinitializes core library classes. Generally these core library classes are read-only and are therefore a good candidate for preloading and sharing across processes. When a new application should be started, Android connects to Zygote through UNIX socket to request the start of a new application. Then Zygote forks, the child process loads the new application in the new instance of the Dalvik VM and start executing it.

# 6 Conclusions

In this paper we first propose a general procedure that can be used on any kind of system to extract generic information flows, highlighting potential problems that can rise during this operation. We then investigate the software architecture of the Android platform, locate different logging infrastructures inside it and show how they work. In this paper we also analyze the Qemu software, that was used by Google as base to develop the Android Emulator. We present its overall architecture, showing the mechanisms that allow this software to emulate a device that is completely different from the machine it is running on. We present Qemu logging infrastructures and how to use them to capture information from an emulated hardware platform. The last part of our work on Qemu is aimed to illustrate the differences between this software and the Android Emulator. In the last part of this paper we analyze what happens inside an Android device on system boot, focusing our attention on differences between Android-based and other Linux-based devices.

This work is aimed to give an overall idea of the different tools that can be used to extract information from Android devices, either physical or emulated. Despite Android open source nature, its internal logging mechanisms are not clearly explained in its documentation. This is because Android's documentation is developer-oriented: it only focus on high-level logging mechanisms, hiding anything not useful to develop Android applications. Similar considerations apply to Qemu: its documentation is often incomplete and not detailed, so that it is necessary to look at the source code or at external resources to fully understand some internal mechanisms. The purpose of this paper is to allow the reader to understand the whole Android and Qemu logging infrastructures without having to dig in the source code or spend days searching the Web and aggregating the gathered information.